\documentclass[a4paper]{jpconf}
\usepackage{graphicx}
\usepackage{xspace}
\usepackage{amsmath}
\usepackage{amssymb}
\usepackage{hyperref}

\newcommand{\pp}{pp\xspace}
\newcommand{\ppb}{p--Pb\xspace}

\newcommand{\ptee}{\ensuremath{p_{\rm{T}}}\xspace}

\newcommand{\dndeta}{\ensuremath{{\rm d}N/{\rm d}\eta}\xspace}
\newcommand{\snnt}[1]{\ensuremath{\sqrt{s_{\rm NN}} = #1 \text{\,TeV}}\xspace}
\newcommand{\snng}[1]{\ensuremath{\sqrt{s_{\rm NN}} = #1 \text{\,GeV}}\xspace}

\newcommand{\gevc}[1]{\ensuremath{#1 \text{\,GeV/$c$}}\xspace}
\newcommand{\raa}{\ensuremath{R_{\rm AA}}\xspace}
\newcommand{\raain}{\ensuremath{R_{\rm AA,in}}\xspace}
\newcommand{\raaout}{\ensuremath{R_{\rm AA,out}}\xspace}

\newcommand{\vtwo}{\ensuremath{v_2}\xspace}
\newcommand{\vtwoo}{\ensuremath{v_{2,a}}\xspace}
\newcommand{\vtwot}{\ensuremath{v_{2,b}}\xspace}
\newcommand{\etwo}{\ensuremath{\varepsilon_2}\xspace}
\newcommand{\etwoa}{\ensuremath{\varepsilon_{2,a}}\xspace}
\newcommand{\etwob}{\ensuremath{\varepsilon_{2,b}}\xspace}
\newcommand{\kf}{\ensuremath{k_{\text{flow}}}\xspace}
\newcommand{\kj}{\ensuremath{k_{\text{medium}}}\xspace}
\newcommand{\kp}{\ensuremath{k_{\text{pathlength}}}\xspace}
\newcommand{\Rf}{\ensuremath{R_{\text{flow}}}\xspace}
\newcommand{\Rj}{\ensuremath{R_{\text{quench}}}\xspace}

\begin{document}
\title{Event-Shape Engineering and Jet Quenching}

\author{Peter Christiansen}

\address{Lund University, Division of Particle Physics, Sweden}

\ead{peter.christiansen@hep.lu.se}

\begin{abstract}
Event-Shape Engineering (ESE) is a tool that enables some control of the
initial geometry in heavy-ion collisions in a similar way as the centrality
enables some control of the number of participants. Utilizing ESE, the path
length in and out-of plane can be varied while keeping the medium properties
(centrality) fixed. In this proceeding it is argued that this provides
additional experimental information about jet quenching. Finally, it is
suggested that if ESE studies are done in parallel for light and heavy quarks
one can determine, in a model independent way, if the path-length dependence
of their quenching differs.
\end{abstract}

\section{Introduction}

Jet quenching offers a possibility to determine properties of the QGP but it
requires that the energy loss mechanism and the effects of the expanding
medium are under control. Results from the LHC have demonstrated that by going
to high \ptee, $\ptee > \gevc{10}$, the \ptee spectra are dominated by
(quenched) jets and so one avoids complicated overlap effects with soft and
intermediate \ptee physics processes. In approximately this \ptee region,
PHENIX has shown~\cite{Adare:2010sp} that energy-loss models in general fails
at describing both the nuclear modification factor, \raa, and the
elliptic-flow coefficient, \vtwo (at high {\ptee}, \vtwo is expected to be
entirely due to jet quenching reflecting the azimuthal asymmetry of the path
length). Recently, some models have been able to describe
both~\cite{Betz:2012qq,Betz:2014cza}, especially for $\ptee >
\gevc{20}$. However, in this author's opinion, it is a complication that a
realistic model must necessarily involve also the initial energy density and
the expansion of the medium, and so one might ask if it is possible to
construct better experimental discriminators. The goal of this proceeding is
to point out that Event-Shape Engineering (ESE) might be such a tool.

This proceeding is outlined as follows. First, the idea is outlined. Second, a
concrete prediction is given for a simple energy-loss-scaling model previously
developed~\cite{Christiansen:2013hya}. Finally, some general ideas are given
on how these results can be extended.

\section{ESE as a jet quenching tool}

The ESE technique~\cite{Schukraft:2012ah} relies on the observation that the
QGP flows like a nearly ideal (reversible) fluid. This means that, as the
initial state ellipticity, $\etwo$, varies event-by-event, the hydrodynamic
flow at low \ptee, $\ptee < \gevc{2}$, will be directly proportional to the
$\etwo$. For a narrow centrality range we therefore have in each event for
$\ptee < \gevc{2}$ that
\begin{equation}
\vtwo(\ptee) = \kf(\ptee) \etwo,
\end{equation}
where $\kf(\ptee)$ is independent of \etwo and derivable from viscous
hydrodynamic modeling of the QGP.

\begin{figure}[htb]
  \begin{center}
   \includegraphics[width=0.3\columnwidth]{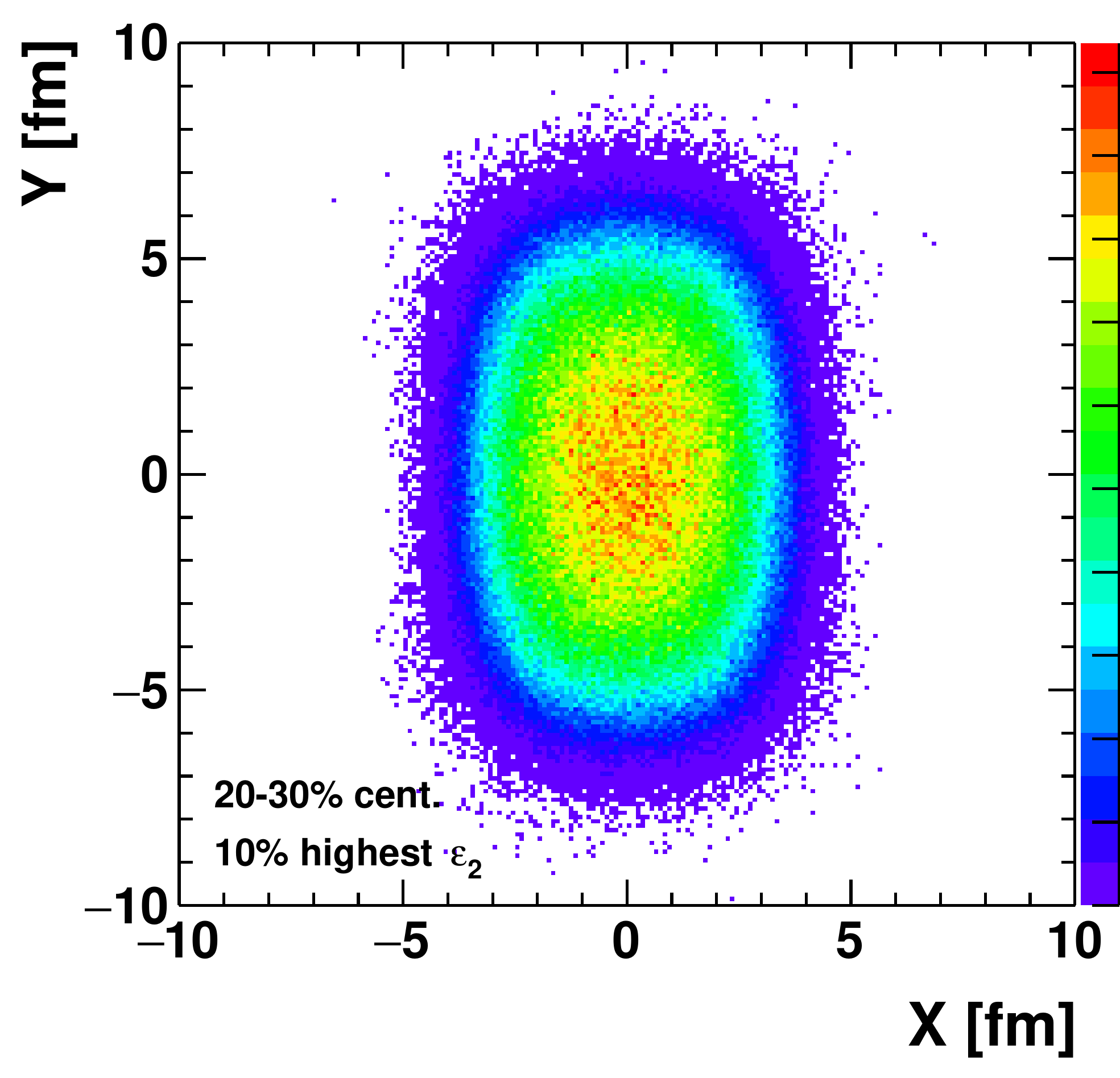}
   \includegraphics[width=0.3\columnwidth]{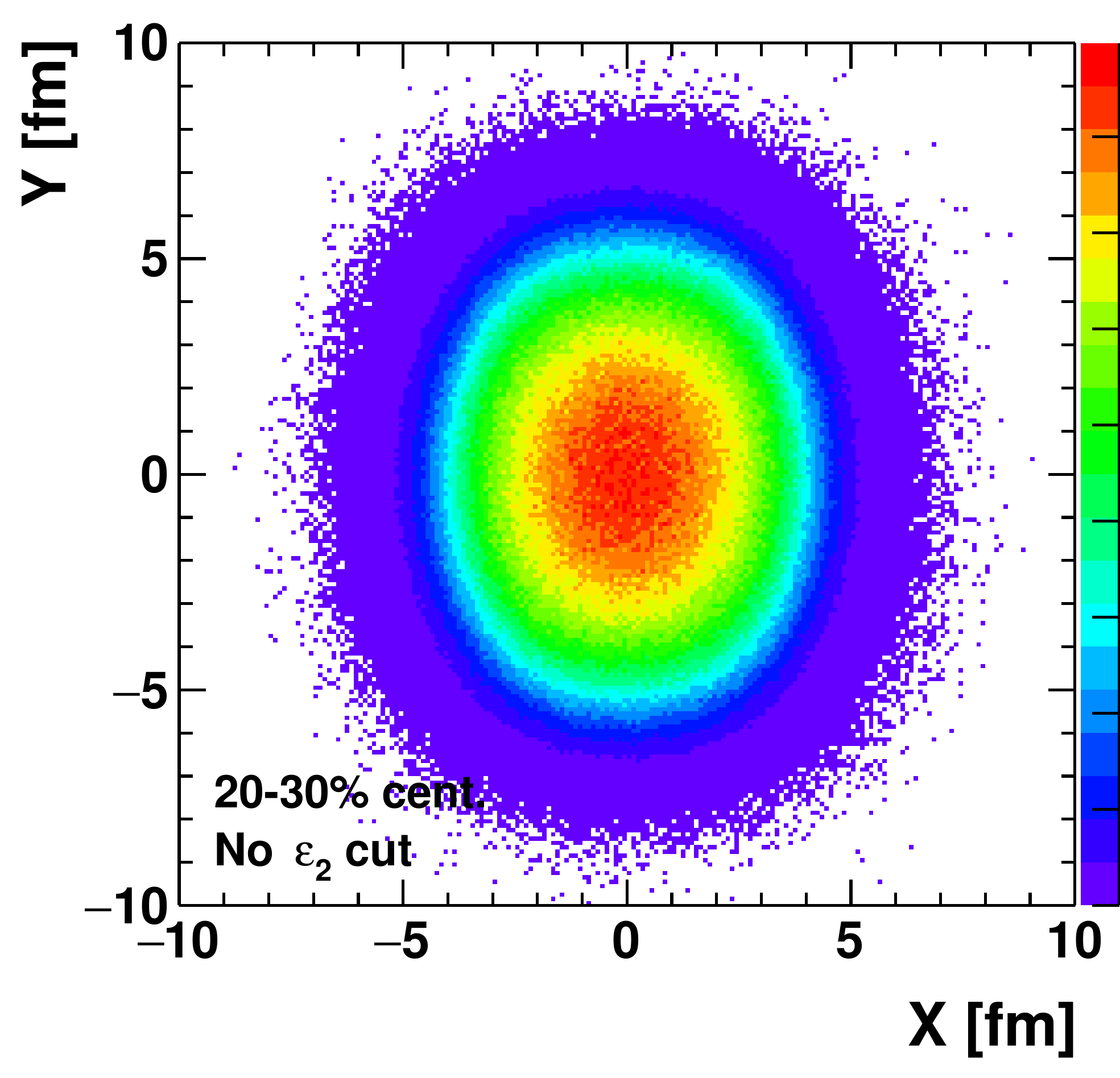}
   \includegraphics[width=0.3\columnwidth]{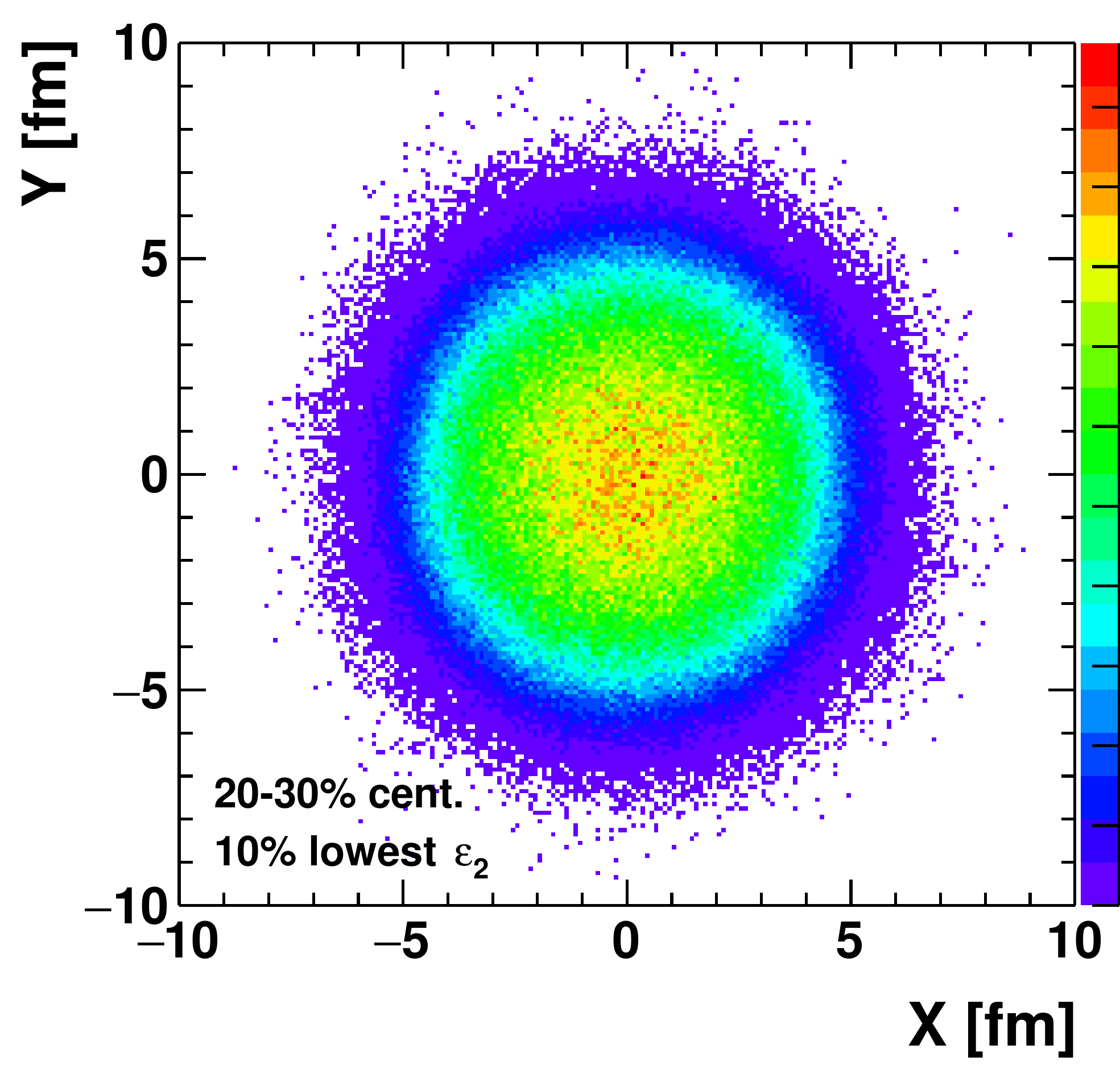}
    \caption{The initial state energy density as given by the $N_{\rm part}$
      distribution for a Glauber calculation of 20-30\% central Pb-Pb
      collisions at \snnt{2.76}. The central plot shows the unbiased
      distribution while the left (right) shows the distribution for the 10\%
      of the events with the highest (lowest) \etwo that can be studied
      using ESE.}
  \label{fig:ese_glauber}
\end{center}
\end{figure}

Figure~\ref{fig:ese_glauber} shows Glauber model calculations for 20-30\%
central Pb-Pb collisions at \snnt{2.76} where in two cases a selection has
been done on \etwo. As can be seen, this changes the geometry dramatically
while keeping the ``area'' of the medium approximately fixed, which suggests
that one does not change the average energy density of the QGP~\footnote{For
  the concrete model in Sec.~\ref{sec:model} the decrease is 10\% going from
  $\etwo \approx 0$ to $\etwo \approx 0.6$. It seems possible to reduce this
  bias further by doing the ESE in narrower centrality bins, see
  e.g.~\cite{Adam:2015eta}.}  but only the path length azimuthal asymmetry. By
cutting on a variable related to the final measured integrated \vtwo (the
length of the 2nd-order flow vector) one can select (ESE) events with these
extreme geometries. Importantly, these ideas have been tested and verified by
experiments~\cite{Aad:2015lwa,Adam:2015eta}.

By being able to vary the path length, $L$, while keeping the medium
properties fixed, e.g. $\hat{q}$, one constrains the model significantly more
than by \raa and \vtwo alone. I hope to convince the reader of this in the
following.

A caveat of many measurements is that detectors have finite resolution and so
selecting the top 10\% \etwo in data will not correspond to the top 10\% in
models. However, by comparing two ESE selections, $a$ and $b$, we can
construct the ratio, \Rf:
\begin{equation}
\Rf \equiv \frac{\langle \vtwoo(\ptee) \rangle}{\langle \vtwot(\ptee) \rangle} (\ptee
< \gevc{2}) = \frac{\kf(\ptee) \langle \etwoa \rangle}{\kf(\ptee) \langle
  \etwob \rangle} = \frac{\langle \etwoa \rangle}{\langle \etwob \rangle},
\end{equation}
which is, to first order, \emph{independent} of \ptee for hydrodynamic flow
(i.e.,\ for $\ptee < \gevc{2}$). We note that this ratio gives us an
experimental measure of the observed geometry variation that can be directly
compared to geometry variations in model calculations, i.e., finite-resolution
effects will make \Rf closer to unity, but that can be mimicked in models by
selecting broader \etwo intervals. In Fig.~\ref{fig:ese_glauber}, $\Rf
=\,1.73$ (0.31) for the highest (lowest) \etwo samples relative to the
unbiased sample.

Now we turn to the high \ptee region and the question of what we expect for the
jet ratio
\begin{equation}
\Rj \equiv \frac{\langle \vtwoo(\ptee) \rangle}{\langle \vtwot(\ptee) \rangle} (\ptee
> \gevc{10}).
\end{equation}
If the jet quenching \vtwo grows faster than the flow \vtwo when increasing
\etwo, then $\Rj > \Rf$ when we compare more asymmetric collisions to unbiased
collisions, but $\Rj < \Rf$ when we compare more symmetric collisions to
unbiased collisions.

The basic understanding of energy loss is that in different limits the
path-length dependence can vary. If interference between adjacent interactions
are important, as is expected for radiation energy loss, then the so-called
Landau-Pomeranchuk-Migdal (LPM) effect will give a quadratic ($L^2$)
pathlength dependence. However, if collisional energy loss plays a large role,
or we are in the Bethe-Heitler (BH) limit where there is no interference, one
expects a linear ($L$) dependence. Finally, in some strongly interacting
QCD-like theories that can be solved via the Maldacena conjecture (the so
called AdS-CFT correspondence) one finds a cubic ($L^3$) path-length
dependence. In the following I will try to explain/justify that the
jet-quenching \vtwo can be approximately factorized like this:
\begin{equation}
\label{eq:Rf}
\vtwo(\ptee) \approx \kj(\ptee) \kp(\etwo),
\end{equation}
where $\kj(\ptee)$ encodes the strength of the coupling to the medium and
$\kp(\etwo)$ encodes the path-length dependence. If we have a light quark and
a gluon propagating through the medium then the gluon will in most models
couple stronger to the QGP than the quark but the path-length dependence of
the energy loss will be the same. Even the energy loss of gluons is therefore
larger, the idea here is that, the relative variation of the energy loss (and
therefore to some approximation \raa and \vtwo) when doing the ESE is the
same. For light and heavy quarks the color factor is the same but there can be
kinematic differences at a fixed \ptee due to the deadcone effect on the
radiative energy loss of heavy quarks and the quark-mass difference for
collisional and radiative energy loss. An example of the resulting energy-loss
differences for light and heavy quarks can be found in Fig.~2
in~\cite{Wicks:2005gt}. But as long as this does not change the path-length
dependence then one would again expect the relative changes to be the
same. However, the factorization will break down if the path-length dependence
varies with \ptee (or in a more positive formulation, one can test this
factorization with experimental data).

If Eq.~\ref{eq:Rf} is a good approximation then:
\begin{equation}
\Rj \approx \frac{\kp(\langle \etwoa \rangle)}{\kp(\langle \etwob
  \rangle)},
\end{equation}
where we expect that $\kp(\etwo)$ is a monotonically increasing function of
$\etwo$.

The claim is therefore that by measuring the ratio of $\vtwo(\ptee)$ at low
and high \ptee for two ESE classes and taking their ratio one can get rid of
the complicated functions $\kf(\ptee)$ and $\kj(\ptee)$ to narrow down the
path-length dependence, $\kp(\etwo)$.

\section{Estimating the ratio \Rj-to-\Rf in a simple model}
\label{sec:model}

In~\cite{Christiansen:2013hya}, a simple model that could describe \raa and
\vtwo at high \ptee was developed. The model describes high-\ptee charged
hadrons with $\ptee \approx \gevc{10}$ based on the following
assumptions/simplifications:
\begin{itemize}
\item The initial geometry is obtained from a Glauber calculation. For each
  event the participating nucleons are centered at (0, 0) and rotated so that
  the 2nd-order symmetry plane $\Psi_2 = 0$, see
  e.g. Fig~\ref{fig:ese_glauber}.
\item Jets are assumed to propagate from (0, 0) and the path length, $L$, is
  calculated as the standard deviation of the source in the direction of
  propagation: $\sigma_X$ in plane and $\sigma_Y$ out-of-plane.
\item The source is assumed to be static and to have a uniform density of
\begin{equation}
\rho = k \frac{\dndeta}{4\pi\sigma_X\sigma_Y},
\end{equation}
where $k$ is an unspecified constant, and $4\pi\sigma_X\sigma_Y$ is an
estimate for the area of the medium. 
\item Only jets propagating in-plane and out-of-plane are considered.
\end{itemize} 

The underlying idea is, just as in the previous section, that jet
quenching depends mainly on two effects: medium density, $\rho$, and path
length, $L$. \\

The \raa and \vtwo can be combined to estimate the \raa in and out-of plane:
\begin{eqnarray}
\label{eq:raain}  \raain(\ptee) & \approx & (1 + 2 \vtwo(\ptee))\raa(\ptee)\\
\label{eq:raaout} \raaout(\ptee) & \approx & (1 - 2 \vtwo(\ptee))\raa(\ptee),
\end{eqnarray}
using results from ALICE~\cite{Abelev:2012hxa} and
ATLAS~\cite{ATLAS:2011ah}. The first step was to attempt to find a scaling
variable $\rho^aL^b$ so that all the \raain and \raaout would follow the same
curve. At first, we were unsure if this was possible because one could expect
that $L$ is not a good variable for the energy loss in plane where the source
is expected to expand. We found that this was possible and took this as an
indication that there is little or no effect on the effective path length of
the transverse expansion, i.e., the path lengths evaluated for the initial
state are meaningful variables for describing jet quenching.

A second issue turned out to be that, once a scaling variable is found the
same scaling variable to any power also works. To find a unique solution, we
demanded that the energy loss is roughly linear in this scaling variable. The
\ptee loss has been estimated in several PHENIX publications,
e.g.~\cite{Adler:2006bw}, assuming that the difference between the expected
and observed \ptee spectra is mainly due to a \ptee shift (loss). If one can
parameterize the \pp spectrum $dN/d\ptee$ by a power law, $k_1 \ptee^{k_2}$,
in the relevant \ptee range, then one obtains a simple expression for the
\ptee loss:
\begin{equation} 
\label{eq:ptloss}
\Delta \ptee = p_{\rm T,0} - p_{\rm T,final} = \left(1-\raa^{-1/(k_2+1)}
\right) p_{\rm T,0},
\end{equation} 
where $p_{\rm T,0}$ is the initial \ptee. This non-linear relation between
energy loss and \raa complicates the intuitive understanding of how the
function $\kp(\etwo)$ should behave because increasing the energy loss
dramatically will eventually lead to much smaller changes in \raa.

In the estimate used here we take into account that quenching not only shifts
but also compresses the spectrum.

\begin{figure}[htb]
  \begin{center}
    \includegraphics[width=0.49\columnwidth]{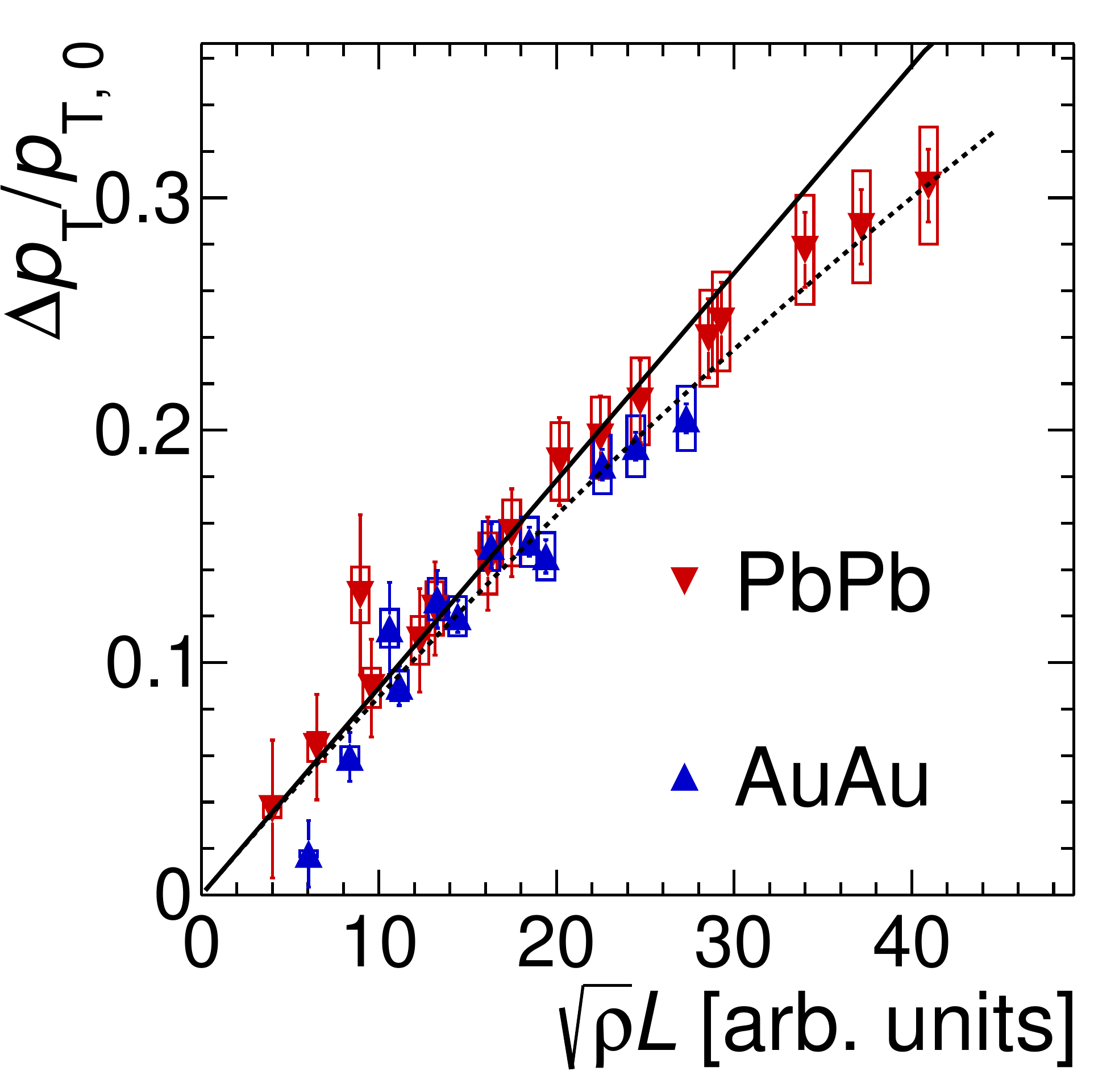}
    \includegraphics[width=0.49\columnwidth]{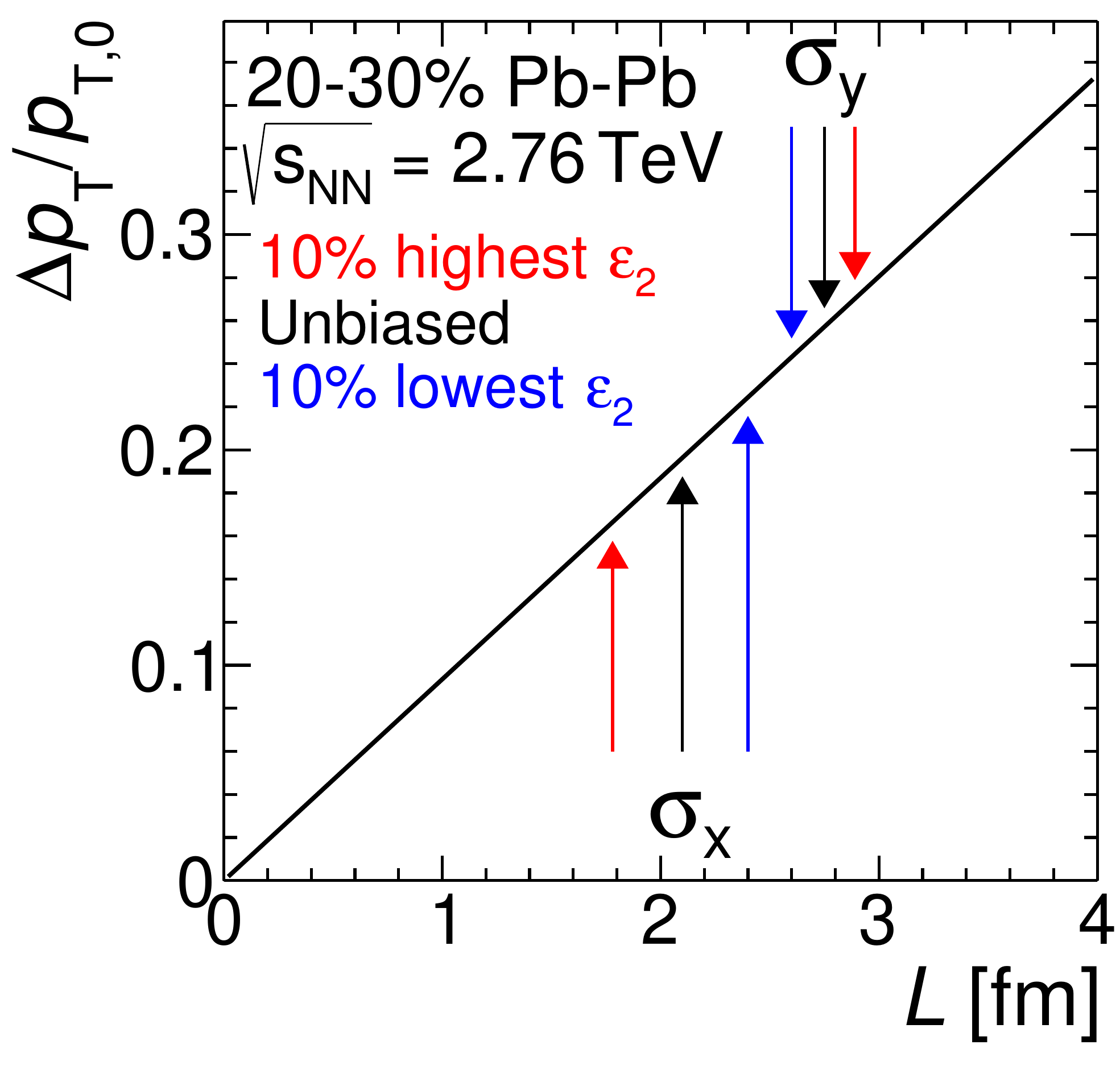}
    \caption{Left: the \ptee loss is found to scale with $\sqrt{\rho}L$ in
      exactly the same way at \snnt{2.76} (PbPb), where the scaling relation
      was derived, and at \snng{200} (AuAu, data taken
      from~\cite{Adare:2012wg}). The solid line is a linear fit. The dashed
      line is a fit that takes into account the decreasing \ptee of the
      propagating particle leading to a small non-linear correction for large
      \ptee losses. Right: the \ptee losses corresponding to the initial
      geometries in Fig.~\ref{fig:ese_glauber} (when the small medium density
      variation is neglected).}
    \label{fig:ptloss}
  \end{center}
\end{figure}

Figure~\ref{fig:ptloss} left shows the main result
of~\cite{Christiansen:2013hya}. The found scaling relation raises numerous
questions that are beyond the scope of this proceeding. Here, I just want to
point out that for symmetric systems, where $L_{\rm in} \approx L_{\rm out}$,
the scaling variable is \emph{independent} of $L$, $\Delta \ptee \propto
\sqrt{\dndeta}$. A similar scaling was recently found by
PHENIX~\cite{Adare:2015cua}. Interestingly, if this result can be trusted,
this would suggest that small systems produced in \pp and \ppb collisions
should also exhibit jet quenching.

The main point of the scaling relation for this proceding is that, once you
have such a model then you can further test it by fixing the medium properties
($\rho$) and varying the path length according to
Fig.~\ref{fig:ese_glauber}. Figure~\ref{fig:ptloss} right shows the predicted
\ptee loss from this model. To obtain \Rj one simply estimates \raain and
\raaout using Eq.~\ref{eq:ptloss} and then determines \vtwo from
Eq.~\ref{eq:raain} and \ref{eq:raaout}. For the concrete model, taking into
account that $\rho$ changes slightly when we vary \etwo, we find:
\begin{eqnarray}
  10\%~\text{highest}~\etwo: & \Rj \approx 1.05 \Rf & \\ 
  10\%~\text{lowest}~\etwo:  & \Rj \approx 0.97 \Rf &
\end{eqnarray}

This means that jet quenching and flow have similar sensitivity to the
azimuthal asymmetry of the initial overlap region.

ATLAS has measured \vtwo as a function of \ptee up to \gevc{15} for different
ESE classes~\cite{ATLAS:2011ah}. Within the large statistical uncertainties
the ratio is flat (Fig.~3 top left in the paper), but for a definitive answer
one will need the higher statistics of Run 2.


Note that recently similar calculations, as have been outlined in this
section, have been done in a realistic model~\cite{Noronha-Hostler:2016eow}.

\section{Conclusions and outlook}

The \raain and \raaout have proven to be difficult observables to describe for
jet-quenching models. By doing these measurements as a function of the
eccentricity one constrains the underlying geometry in a way that simplifies
the direct interpretation of the results and that allows further tests of
models.

Still, one does not avoid the need of a model to interpret the data. To be
able to avoid models one would need two different probes. If one would measure
\Rj for light \emph{and} heavy quarks then one could in principle compare
these directly to understand if the effective path-length dependence of the
energy loss is different, as expected from theoretical models. There is no
reason to expect that the jet quenching \vtwo of light and heavy quarks are
the same because the coupling to the medium, $\kj(\ptee)$, can be different,
but if the path-length dependence, $\kp(\ptee)$, is the same then one would to
first order expect that \Rj is the same (due to the non-linearity of energy
loss and \raa, small differences can be expected) and so one can in a model
independent way test if the path-length dependence, which is directly related
to the physics mechanism of the energy loss, is the same or not.

\ack

The author would like to thank the organizers for a wonderful conference,
Vytautas Vislavicius for making the additional Glauber-model calculations for
these studies, and Jamie Nagle for comments to the proceeding.

\section*{References}

\bibliographystyle{utphys}
\bibliography{biblio}

\providecommand{\href}[2]{#2}\begingroup\raggedright\begin{thebibliography}{10}

\bibitem{Adare:2010sp}
{\bfseries PHENIX} Collaboration, A.~Adare {\em et~al.}, ``{Azimuthal
  anisotropy of neutral pion production in Au+Au collisions at $\sqrt{s_{\rm
  NN}}$ = 200 GeV: Path-length dependence of jet quenching and the role of
  initial geometry},''
  \href{http://dx.doi.org/10.1103/PhysRevLett.105.142301}{{\em Phys. Rev.
  Lett.} {\bfseries 105} (2010) 142301},
\href{http://arxiv.org/abs/1006.3740}{{\ttfamily arXiv:1006.3740 [nucl-ex]}}.

\bibitem{Betz:2012qq}
B.~Betz and M.~Gyulassy, ``{Examining a reduced jet-medium coupling in Pb+Pb
  collisions at the Large Hadron Collider},''
  \href{http://dx.doi.org/10.1103/PhysRevC.86.024903}{{\em Phys. Rev. C}
  {\bfseries 86} (2012) 024903},
\href{http://arxiv.org/abs/1201.0281}{{\ttfamily arXiv:1201.0281 [nucl-th]}}.

\bibitem{Betz:2014cza}
B.~Betz and M.~Gyulassy, ``{Constraints on the Path-Length Dependence of Jet
  Quenching in Nuclear Collisions at RHIC and LHC},''
  \href{http://dx.doi.org/10.1007/JHEP10(2014)043,
  10.1007/JHEP08(2014)090}{{\em JHEP} {\bfseries 08} (2014) 090},
  \href{http://arxiv.org/abs/1404.6378}{{\ttfamily arXiv:1404.6378 [hep-ph]}}.
[Erratum: JHEP10,043(2014)].

\bibitem{Christiansen:2013hya}
P.~Christiansen, K.~Tywoniuk, and V.~Vislavicius, ``{Universal scaling
  dependence of QCD energy loss from data driven studies},''
  \href{http://dx.doi.org/10.1103/PhysRevC.89.034912}{{\em Phys. Rev. C}
  {\bfseries 89} (2014) 034912},
\href{http://arxiv.org/abs/1311.1173}{{\ttfamily arXiv:1311.1173 [hep-ph]}}.

\bibitem{Schukraft:2012ah}
J.~Schukraft, A.~Timmins, and S.~A. Voloshin, ``{Ultra-relativistic nuclear
  collisions: event shape engineering},''
  \href{http://dx.doi.org/10.1016/j.physletb.2013.01.045}{{\em Phys. Lett. B}
  {\bfseries 719} (2013) 394--398},
\href{http://arxiv.org/abs/1208.4563}{{\ttfamily arXiv:1208.4563 [nucl-ex]}}.

\bibitem{Adam:2015eta}
{\bfseries ALICE} Collaboration, J.~Adam {\em et~al.}, ``{Event shape
  engineering for inclusive spectra and elliptic flow in Pb-Pb collisions at
  $\sqrt{s_{\rm NN}}=2.76$ TeV},''
  \href{http://dx.doi.org/10.1103/PhysRevC.93.034916}{{\em Phys. Rev. C}
  {\bfseries 93} (2016) 034916},
\href{http://arxiv.org/abs/1507.06194}{{\ttfamily arXiv:1507.06194 [nucl-ex]}}.

\bibitem{Aad:2015lwa}
{\bfseries ATLAS} Collaboration, G.~Aad {\em et~al.}, ``{Measurement of the
  correlation between flow harmonics of different order in lead-lead collisions
  at $\sqrt{s_{\rm NN}}$=2.76 TeV with the ATLAS detector},''
  \href{http://dx.doi.org/10.1103/PhysRevC.92.034903}{{\em Phys. Rev. C}
  {\bfseries 92} (2015) 034903},
\href{http://arxiv.org/abs/1504.01289}{{\ttfamily arXiv:1504.01289 [hep-ex]}}.

\bibitem{Wicks:2005gt}
S.~Wicks, W.~Horowitz, M.~Djordjevic, and M.~Gyulassy, ``{Elastic, inelastic,
  and path length fluctuations in jet tomography},''
  \href{http://dx.doi.org/10.1016/j.nuclphysa.2006.12.048}{{\em Nucl. Phys.}
  {\bfseries A784} (2007) 426--442},
\href{http://arxiv.org/abs/nucl-th/0512076}{{\ttfamily arXiv:nucl-th/0512076
  [nucl-th]}}.

\bibitem{Abelev:2012hxa}
{\bfseries ALICE} Collaboration, B.~Abelev {\em et~al.}, ``{Centrality
  Dependence of Charged Particle Production at Large Transverse Momentum in
  Pb--Pb Collisions at $\sqrt{s_{\rm{NN}}} = 2.76$ TeV},''
  \href{http://dx.doi.org/10.1016/j.physletb.2013.01.051}{{\em Phys. Lett. B}
  {\bfseries 720} (2013) 52--62},
\href{http://arxiv.org/abs/1208.2711}{{\ttfamily arXiv:1208.2711 [hep-ex]}}.

\bibitem{ATLAS:2011ah}
{\bfseries ATLAS} Collaboration, G.~Aad {\em et~al.}, ``{Measurement of the
  pseudorapidity and transverse momentum dependence of the elliptic flow of
  charged particles in lead-lead collisions at $\sqrt{s_{NN}}=2.76$ TeV with
  the ATLAS detector},''
  \href{http://dx.doi.org/10.1016/j.physletb.2011.12.056}{{\em Phys. Lett. B}
  {\bfseries 707} (2012) 330--348},
\href{http://arxiv.org/abs/1108.6018}{{\ttfamily arXiv:1108.6018 [hep-ex]}}.

\bibitem{Adler:2006bw}
{\bfseries PHENIX} Collaboration, S.~S. Adler {\em et~al.}, ``{A Detailed Study
  of High-p(T) Neutral Pion Suppression and Azimuthal Anisotropy in Au+Au
  Collisions at s(NN)**(1/2) = 200-GeV},''
  \href{http://dx.doi.org/10.1103/PhysRevC.76.034904}{{\em Phys. Rev. C}
  {\bfseries 76} (2007) 034904},
\href{http://arxiv.org/abs/nucl-ex/0611007}{{\ttfamily arXiv:nucl-ex/0611007
  [nucl-ex]}}.

\bibitem{Adare:2012wg}
{\bfseries PHENIX} Collaboration, A.~Adare {\em et~al.}, ``{Neutral pion
  production with respect to centrality and reaction plane in Au$+$Au
  collisions at $\sqrt{s_{NN}}$=200 GeV},''
  \href{http://dx.doi.org/10.1103/PhysRevC.87.034911}{{\em Phys. Rev. C}
  {\bfseries 87} (2013) 034911},
\href{http://arxiv.org/abs/1208.2254}{{\ttfamily arXiv:1208.2254 [nucl-ex]}}.

\bibitem{Adare:2015cua}
{\bfseries PHENIX} Collaboration, A.~Adare {\em et~al.}, ``{Scaling properties
  of fractional momentum loss of high-$p_T$ hadrons in nucleus-nucleus
  collisions at $\sqrt{s_{_{NN}}}$ from 62.4 GeV to 2.76 TeV},''
  \href{http://dx.doi.org/10.1103/PhysRevC.93.024911}{{\em Phys. Rev.}
  {\bfseries C93} (2016) 024911},
\href{http://arxiv.org/abs/1509.06735}{{\ttfamily arXiv:1509.06735 [nucl-ex]}}.

\bibitem{Noronha-Hostler:2016eow}
J.~Noronha-Hostler, B.~Betz, J.~Noronha, and M.~Gyulassy, ``{Soft-Hard Event
  Engineering in Ultrarelativistic Heavy Ion Collisions},''
\href{http://arxiv.org/abs/1602.03788}{{\ttfamily arXiv:1602.03788 [nucl-th]}}.

\end{thebibliography}\endgroup

\end{document}